\documentclass[12pt]{article}
\usepackage{graphicx}
\usepackage{epsf}
\def\beq{\begin{equation}}
\def\eeq{\end{equation}}
\def\bea{\begin{eqnarray}}
\def\eea{\end{eqnarray}}
\def\nn{\nonumber}
\def\sss{\scriptscriptstyle}
\def\roughly#1{\mathrel{\raise.3ex\hbox
{$#1$\kern-.75em\lower1ex\hbox{$\sim$}}}}

\def\gsim{\roughly>}
\def\bd{B_d^0}
\def\bs{B_s^0}
\def\bdbar{{\bar B}_d^0}
\def\bsbar{{\bar B}_s^0}

\def\btos{{\bar b} \to {\bar s}}

\def\kbar{{\overline{K^0}}}

\def\pewp{P'_{\sss EW}}

\def\btopik{B \to \pi K}
\def\bra#1{\left\langle #1\right|}
\def\ket#1{\left| #1\right\rangle}
\def\gf{G_{\sss F}}
\def\wt{\widetilde}
\def\ANPq{{\cal A}^q}

\def\ApNPcom{{\cal A}^{\prime, comb} }
\def\ApNPqph{{\cal A}^{\prime,q} e^{i \Phi'_q}}
\def\ApNPCqph{{\cal A}^{\prime {\sss C}, q} e^{i \Phi_q^{\prime C}}}
\def\ApNPCuph{{\cal A}^{\prime {\sss C}, u} e^{i \Phi_u^{\prime C}}}
\def\ApNPCdph{{\cal A}^{\prime {\sss C}, d} e^{i \Phi_d^{\prime C}}}
\def\ApNPuph{{\cal A}^{\prime,u} e^{i \Phi'_u}}
\def\ApNPdph{{\cal A}^{\prime,d} e^{i \Phi'_d}}
\def\ApNPcomb{{\cal A}^{\prime, comb} e^{i \Phi'}}
\def\bskk{{\bs}\to K^+ K^-}
\def\bskkneut{{\bs}\to K^0 \kbar}
\def\ba{\begin{array}}
\def\ea{\end{array}}
\def\l{\left}
\def\r{\right}
\def\te{\theta}
\def\de{\delta}

\begin{document}

\begin{flushright}
UdeM-GPP-TH-07-164 \\
\end{flushright}

\begin{center}
\bigskip
{\Large \bf \boldmath The $\btopik$ Puzzle and Supersymmetry}
\\
\bigskip
\bigskip
Maxime Imbeault
$^{a,}$\footnote{maxime.imbeault@umontreal.ca},
Seungwon Baek $^{b,}$\footnote{swbaek@kias.re.kr},
and David London $^{a,}$\footnote{london@lps.umontreal.ca}
\end{center}
\begin{flushleft}
~~~~~~~~~~~$a$: {\it Physique des Particules, Universit\'e
de Montr\'eal,}\\
~~~~~~~~~~~~~~~{\it C.P. 6128, succ. centre-ville,
Montr\'eal, QC, Canada H3C 3J7}\\
~~~~~~~~~~~$b$: {\it The Institute of Basic Science and
Department of Physics,}\\
~~~~~~~~~~~~~~~{\it Korea University, Seoul 136-701,
    Korea. }\\
\end{flushleft}

\begin{center}
\bigskip (\today)
\vskip0.5cm {\Large Abstract\\} \vskip3truemm
\parbox[t]{\textwidth}{At present, there are discrepancies
between the measurements of several observables in $\btopik$
decays and the predictions of the standard model (the
``$\btopik$ puzzle'').  Although the effect is not yet
statistically significant -- it is at the level of $\gsim
3\sigma$ -- it does hint at the presence of new physics. In
this paper, we explore whether supersymmetry (SUSY) can
explain the $\btopik$ puzzle.  In particular, we consider the
SUSY model of Grossman, Neubert and Kagan (GNK).  We find
that it is extremely unlikely that GNK explains the $\btopik$
data.  We also find a similar conclusion in many other models
of SUSY.  And there are serious criticisms of the two SUSY
models that do reproduce the $\btopik$ data. If the $\btopik$
puzzle remains, it could pose a problem for SUSY models.}
\end{center}

\thispagestyle{empty}
\newpage
\setcounter{page}{1}
\baselineskip=14pt

Over the past several years, measurements have been made of a
number of observables in the decays of $B$ mesons which are
in disagreement with the predictions of the standard model
(SM): e.g.\ indirect CP asymmetries in penguin-dominated $B$
decays \cite{HFAG}, triple-product correlations in $B \to
\phi K^*$ \cite{BaBarTP}, polarizations in $B\to V_1V_2$
decays ($V_i$ is a vector meson) \cite{polarization},
etc. None of these discrepancies is statistically
significant, so that these disagreements only point to a hint
of physics beyond the SM. Still, if these hints are taken
together, the statistical significance increases.
Furthermore, they are intriguing since they all point to new
physics (NP) in $\btos$ transitions.

Arguably, the most stringent discrepancy appears in $\btopik$
decays.  Briefly, the effect goes as follows.  There are four
$\btopik$ decays: $B^+ \to \pi^+ K^0$ (designated as $+0$
below), $B^+ \to \pi^0 K^+$ ($0+$), $\bd \to \pi^- K^+$
($-+$) and $\bd \to \pi^0 K^0$ ($00$). In terms of diagrams
\cite{GHLR}, the amplitudes are given by
\bea
\label{fulldiagrams}
A^{+0} & = & -P' ~, \nn\\
\sqrt{2} A^{0+} & = & P' -T' e^{i\gamma} -C'
e^{i\gamma} - \pewp ~, \nn\\
A^{-+} & = & P' -T' e^{i\gamma} ~, \nn\\
\sqrt{2} A^{00} & = & -P' - \pewp -C' e^{i\gamma} ~.
\label{piKampsSM}
\eea
In the above, we have neglected small diagrams and written
the amplitudes in terms of the color-favored and
color-suppressed tree amplitudes $T'$ and $C'$, the
$t$-quark-dominated gluonic penguin amplitude $P'$, and the
color-favored electroweak penguin amplitude $\pewp$. (The
primes on the amplitudes indicate $\btos$ transitions.) In
addition, we have explicitly written the weak-phase
dependence (including the minus sign from $V_{tb}^* V_{ts}$
[$P'$]), while the diagrams contain strong phases. (The phase
information in the Cabibbo-Kobayashi-Maskawa (CKM) quark
mixing matrix is conventionally parametrized in terms of the
unitarity triangle, in which the interior (CP-violating)
angles are known as $\alpha$, $\beta$ and $\gamma$
\cite{pdg}.) The amplitudes for the CP-conjugate processes
can be obtained from the above by changing the sign of the
weak phase ($\gamma$). Note that these diagrams include the
magnitudes of their associated CKM matrix elements.

The diagram $\pewp$ is not independent.  To a good
approximation, it can be related to $T'$ and $C'$ using
flavor SU(3) symmetry \cite{EWPs}:
\beq
\label{EWPrels}
\pewp = {3\over 4} {c_9 + c_{10} \over c_1 + c_2} R (T' + C')
\!+\!  {3\over 4} {c_9 - c_{10} \over c_1 - c_2} R (T' - C')
~.
\eeq
Here, the $c_i$ are Wilson coefficients \cite{BuraseffH} and
$R \equiv \left\vert (V_{tb}^* V_{ts})/(V_{ub}^* V_{us})
\right\vert$.

Now, in Ref.~\cite{GHLR}, the relative sizes of the diagrams
were estimated to be roughly
\beq
1 : |P'| ~~,~~~~ {\cal O}({\bar\lambda}) : |T'|,~|\pewp|
~~,~~~~ {\cal O}({\bar\lambda}^2) : |C'| ~.
\label{hierarchy}
\eeq
where ${\bar\lambda} \sim 0.2$. With this estimate, the
diagram $C'$ should also be neglected in the $\btopik$
amplitudes above [Eq.~(\ref{piKampsSM})]. Note that the
smallness of $|C'|$ is verified by more robust hadronic
computations: $|C'/T'| \sim 0.3$ is the prediction of NLO
pQCD \cite{PQCD_NLO}, and $|C'/T'| \sim 0.6$ is the maximal
SCET (QCDf) prediction \cite{Beneke:2003zv,SCET}.

There are nine measurements that have been made of $\btopik$
decays: the four branching ratios, the four direct CP
asymmetries $A_{\sss CP}^{ij}$ ($ij = +0$, $0+$, $-+$, $00$),
and the mixing-induced CP asymmetry $S_{\sss CP}^{00}$ in
$\bd\to \pi^0K^0$ \cite{piKrefs}. With this data and the
expressions for the $\btopik$ amplitudes, one can perform a
fit \cite{BtopiKfits}.  In the first fit, $C'$ was neglected
in the $\btopik$ amplitudes. A very poor fit was found:
$\chi^2_{min}/d.o.f. = 25.0/5~(1.4\times 10^{-4})$. (The
number in parentheses indicates the quality of the fit, and
depends on $\chi^2_{min}$ and $d.o.f.$ individually.  It
shows the percentage of the parameter space which has a worse
$\chi^2_{min}$.  50\% or more is a very good fit; fits which
are substantially less than 50\% are poorer. $1.4\times
10^{-4}$ corresponds to a 3-4$\sigma$ discrepancy with the
SM.) This result has led some authors to posit the existence
of a ``$\btopik$ puzzle'' \cite{BFRS}.

In the second fit, $C'$ was kept and the full amplitudes of
Eq.~(\ref{fulldiagrams}) used. In this case, a good fit was
found: $\chi^2_{min}/d.o.f. =1.0/3~(80\%)$. This has led some
people to argue that there is in fact no $\btopik$ puzzle
(for example, see Ref.~\cite{CKMfitterpub}). However,
$|C'/T'| = 1.6 \pm 0.3$ is required here. This is much larger
then the theoretical estimates described above.  If one takes
this theoretical input seriously -- as we do here -- this
shows explicitly that the $\btopik$ puzzle is still present,
at $\gsim$ the $3\sigma$ level.

The question now is: what type of new physics can explain the
$\btopik$ puzzle? All NP operators in $\btos q {\bar q}$
transitions take the form ${\cal O}_{\sss NP}^{ij,q} \sim
{\bar s} \Gamma_i b \, {\bar q} \Gamma_j q$ ($q = u,d,s,c$),
where the $\Gamma_{i,j}$ represent Lorentz structures, and
color indices are suppressed. These operators contribute to
the decay $\btopik$ through the matrix elements $\bra{\pi K}
{\cal O}_{\sss NP}^{ij,q} \ket{B}$. Each matrix element has
its own NP weak and strong phase. Now, it has been argued
that all NP strong phases are negligible \cite{DLNP}. In this
case one can combine all NP matrix elements of $\btopik$ into
a single NP amplitude, with a single weak phase:
\beq
\sum \bra{\pi K} {\cal O}_{\sss NP}^{ij,q} \ket{B} = \ANPq
e^{i \Phi_q} ~.
\eeq
$\btopik$ decays involve only NP parameters related to the
quarks $u$ and $d$. These operators come in two classes,
differing in their color structure: ${\bar s}_\alpha \Gamma_i
b_\alpha \, {\bar q}_\beta \Gamma_j q_\beta$ and ${\bar
s}_\alpha \Gamma_i b_\beta \, {\bar q}_\beta \Gamma_j
q_\alpha$ ($q=u,d$). The matrix elements of these operators
can be combined into single NP amplitudes, denoted $\ApNPqph$
and $\ApNPCqph$, respectively \cite{BNPmethods}. Here,
$\Phi'_q$ and $\Phi_q^{\prime {\sss C}}$ are the NP weak
phases; the strong phases are zero. Each of these contributes
differently to the various $\btopik$ decays. In general,
${\cal A}^{\prime,q} \ne {\cal A}^{\prime {\sss C}, q}$ and
$\Phi'_q \ne \Phi_q^{\prime {\sss C}}$. Note that, despite
the ``color-suppressed'' index $C$, the matrix elements
$\ApNPCqph$ are not necessarily smaller than the $\ApNPqph$.

The $\btopik$ amplitudes can now be written in terms of the
SM amplitudes to $O({\bar\lambda})$ [$\pewp$ and $T'$ are
related as in Eq.~(\ref{EWPrels})], along with the NP matrix
elements \cite{BNPmethods}:
\bea
\label{BpiKNPamps}
A^{+0} & = & -P' + \ApNPCdph ~, \\
\sqrt{2} A^{0+} & = & P' - T' \, e^{i\gamma} +
\pewp +~\ApNPcomb - \ApNPCuph ~, \nn\\
A^{-+} & = & P' - T' \, e^{i\gamma} - \ApNPCuph
~, \nn\\
\sqrt{2} A^{00} & = & -P' + \pewp
+ \ApNPcomb + \ApNPCdph ~, \nn
\eea
where $\ApNPcomb \equiv - \ApNPuph + \ApNPdph$.

In 1999, Grossman, Neubert and Kagan (GNK) proposed a new
version of supersymmetry (SUSY) \cite{GNK}. This model was
promising for NP contributions to $\btopik$ decays because it
incorporates a new CP phase, and because it breaks isospin.
In this paper we explore whether the GNK SUSY model can in
fact explain the $\btopik$ puzzle, i.e.\ whether it gives the
appropriate contributions to $\ApNPcomb$, $\ApNPCuph$ and
$\ApNPCdph$.

We begin with a review of the GNK SUSY model, emphasizing
those points which are important to our calculation. In
R-parity-conserving SUSY models, the largest contributions to
flavor-changing neutral current (FCNC) processes potentially
come from the gluino-exchange SUSY box or penguin
diagrams. The chargino and neutralino contributions are
parametrically suppressed due to their small gauge
couplings. The source of the gluino-mediated FCNC is the
off-diagonal components in the scalar mass matrix in the
basis where the quark mass matrices are diagonalized
(super-CKM basis).  Since we are interested only in the
$\btos$ transition, we consider only the down-type scalar
mass matrix.

However, a generic form of scalar mass matrices is not
acceptable because it leads to too-large contributions to
FCNC processes (SUSY FCNC problem) and/or to the electric
dipole moments of the neutron and electron (SUSY CP
problem). To evade these problems, people usually assume that
SUSY is broken in a hidden sector and mediated to the
observable sector by some flavor-blind interactions, such as
gravity or gauge interactions. Then the squark mass matrices
are diagonal matrices at a high-energy scale. The
off-diagonal components in the squark mass matrices are
generated by renormalization group (RG) running. In these
popular models, such as minimal supergravity (mSUGRA)
\cite{mSUGRA}, anomaly-mediated SUSY breaking (AMSB)
\cite{AMSB} or gauge-mediated SUSY breaking (GMSB)
\cite{GMSB} models, the SUSY FCNC/CP problems are solved
because the RG-generated off-diagonal terms are typically
very small and they do not include new sources of CP
violation. On the other hand, as a consequence, they also
cannot explain any possible deviation in the CP asymmetries
in $B$ decays.

The GNK model assumes the following form of sdown
mass-squared matrices:
\bea
 M^2_{\sss \wt{d},LL(RR)}
=\left(\begin{array}{ccc}
 \wt{m}^{\sss d,2}_{\sss L(R)_{11}} & 0 & 0 \\
 0 & \wt{m}^{\sss d,2}_{\sss L(R)_{22}} & \wt{m}^{\sss d,2}_{\sss L(R)_{23}} \\
 0 & \wt{m}^{\sss d,2}_{\sss L(R)_{32}} & \wt{m}^{\sss d,2}_{\sss L(R)_{33}} \\
\end{array}
\right), \quad
 M^2_{\sss \wt{d},LR(RL)} \equiv 0_{3\times 3} ~,
\label{eq:sdown}
\eea
where off-diagonal components can be as large as the diagonal
components.  Although Eq.~(\ref{eq:sdown}) is not supported
by the above-mentioned popular SUSY-breaking models, it is
well-motivated in SUSY GUT theories, where neutrinos are in
the same supermultiplet as down quarks \cite{SUSYGUT}.  The
zeroes in the above mass matrix are justified by the fact
that the experimental results for $K^0$-$\bar{K}^0$ mixing,
$\bd$-$\bdbar$ mixing and $B \to X_s \gamma$ are in good
agreement with the SM predictions. In general they can get
small non-zero values, but they do not affect our results
much as long as we do not consider the very large $\tan
\beta$ region~\cite{double_MI}. In our analysis below, we
consider two scenarios: (i) only $LL$ mixing is present
(i.e.\ $M^2_{\sss \wt{d},RR}$ is diagonal), and (ii) both
$LL$ and $RR$ mixing are present.

The mass matrix $M^2_{\sss \wt{d},LL}$ is diagonalized by
\bea
 \Gamma_{\sss L} M^2_{\sss \wt{d},LL} \Gamma_{\sss L}^\dagger
= {\rm diag}(m^2_{\sss \wt{d}_{\sss L}},m^2_{\sss
 \wt{s}_{\sss L}},m^2_{\sss \wt{b}_{\sss L}}) ~,
\eea
with
\bea
 \Gamma_{\sss L}
=\l(\ba{ccc}
 1 & 0 & 0 \\
 0 & \cos\te_{\sss L} & \sin\te_{\sss L}\;e^{i\de_{\sss L}} \\
 0 & -\sin\te_{\sss L}\;e^{-i\de_{\sss L}} & \cos\te_{\sss L}  \\
\ea
\r) ~.
\label{eq:Gamma_L}
\eea
Similarly, the exchange $L \leftrightarrow R$ in
(\ref{eq:Gamma_L}) gives $\Gamma_{\sss R}$.  We restrict to
$-\pi/4 < \te_{\sss L(R)} < \pi/4$ ($\te_{\sss R} = 0$ if
$RR$ mixing is absent) and $-\pi < \de_{\sss L(R)} < \pi$.

The form given in Eq.~(\ref{eq:sdown}) is not sufficient to
give large SUSY contributions to $\pewp$. (Actually it is
known that the gluino contribution to the $Z$-penguin is
small \cite{Nir:1997tf}.)  In Ref.~\cite{GNK}, the authors
assumed that there is a significant mass splitting between
the right-handed up and down squarks. Then the gluino box
diagrams become the main source of the isospin breaking, and
the scale $\alpha_s^2/m_{\sss SUSY}^2$ (SUSY contribution) is
comparable with $\alpha/M_{\sss W}^2$ (SM contribution)

We now turn to a review of the new-physics amplitudes
$\ApNPcomb$, $\ApNPCuph$ and $\ApNPCdph$. These same NP
amplitudes also contribute to $\bskk$ and $\bskkneut$, and
have been calculated within GNK SUSY in Ref.~\cite{BsKK}.  We
closely follow this reference in our analysis, and use its
treatment of the NP SUSY amplitudes. The color-allowed and
color-suppressed NP amplitudes are given by
\bea
\ApNPqph &=& \frac{\gf}{\sqrt{2}} \left[ (\bar c_1^q +
\frac{1}{3}\bar c_2^q) - (\bar c_3^q + \frac{1}{3} c_4^q) -
\chi_\pi (\frac{1}{3}\bar c_5^q + \bar c_6^q) \right]
A_{\sss K \pi}~, \nn\\
\ApNPCqph &=& \frac{\gf}{\sqrt{2}}\left[ -\chi_{\sss K}
(\frac{1}{3}\bar c_1^q + \bar c_2^q ) - (\frac{1}{3} \bar
c_3^q + \bar c_4^q) + (\bar c_5^q + \frac{1}{3} \bar c_6^q)
\right. \nn\\
& & ~~~~~~~~~~~~~~~~~~~~~~~~~~~~~ \left.
- \lambda_t \frac{2 \alpha_s}{3 \pi} \bar c_{8g}^{eff}
(1+\frac{\chi_{\sss K}}{3}) \right]A_{\sss\pi K},
\label{EqACA}
\eea
where $q=u,d$. (Note: in Ref.~\cite{BsKK}, $\ApNPCqph$ and
$\ApNPqph$ are switched.)  In the above, $\lambda_t =
V_{tb}^* V_{ts}$ and
\bea
{\bar c}_i^q &=& c_i - \tilde c_i~,\nn\\
{\bar c}_{8g}^{eff} &=& c_{8g} +
\frac{c_1^u+2c_1^d}{3}~,\nn\\
A_{\sss \pi K} &=& i (m_{\sss B}^2 -
m_\pi^2) F_0^{\sss B\to \pi} (m_{\sss K}^2) f_{\sss K}~,\nn\\
A_{\sss K \pi} &=& i (m_{\sss B}^2 -
m_{\sss K}^2) F_0^{\sss B\to K} (m_\pi^2) f_\pi~,\nn\\
\chi_{\sss K} (\mu ) &=& \frac{2 m_{\sss K}^2}{\bar m_b (\mu
) (\bar m_q (\mu) + \bar m_s (\mu))}~,\nn\\
\chi_\pi (\mu ) &=& \frac{2 m_\pi^2}{\bar m_b (\mu ) (\bar
m_u (\mu) + \bar m_d (\mu))}~,
\eea
where the $c$'s and $\tilde c$'s are Wilson coefficients of
the effective operator in the GNK basis, $m_q$ is the
averaged mass of up and down quarks, and naive factorization
has been used for the hadronic matrix elements $A_{\sss \pi
K}$ and $A_{\sss K \pi}$. Also, $c_{8g} = -\lambda_t C_{8g}$,
where $C_{8g}^{eff} = C_{8g} + C_5$ in the standard basis.

When only mixing between components 2 and 3 of the
down-squark mixing matrices is allowed, the Wilson
coefficients are given by
\bea
c_1^q &=& \frac{\alpha_s^2 \sin{2 \theta_{\sss L}} e^{i
\delta_{\sss L}}}{4 \sqrt{2} \gf m_{\tilde g}^2} \left[
\frac{1}{18}F(x_{\tilde b_{\sss L} \tilde g},x_{\tilde q_R
\tilde g}) -\frac{5}{18} G(x_{\tilde b_{\sss L} \tilde
g},x_{\tilde q_R \tilde g}) + \frac{1}{2} A(x_{\tilde b_{\sss
L} \tilde g}) + \frac{2}{9} B(x_{\tilde b_{\sss L} \tilde
g})\right] \nn\\
& & ~~~~~~~~~~~~~~~~~~~~
- (\tilde b_{\sss L} \to \tilde s_{\sss L})~,\nn\\
c_2^q &=& \frac{\alpha_s^2 \sin{2 \theta_{\sss L}} e^{i
\delta_{\sss L}}}{4 \sqrt{2} \gf m_{\tilde g}^2} \left[
\frac{7}{6}F(x_{\tilde b_{\sss L} \tilde g},x_{\tilde q_R
\tilde g}) +\frac{1}{6} G(x_{\tilde b_{\sss L} \tilde
g},x_{\tilde q_R \tilde g}) - \frac{3}{2} A(x_{\tilde b_{\sss
L} \tilde g}) - \frac{2}{3} B(x_{\tilde b_{\sss L} \tilde
g})\right] \nn\\
& & ~~~~~~~~~~~~~~~~~~~~
- (\tilde b_{\sss L} \to \tilde s_{\sss L})~,\nn\\
c_3^q &=& \frac{\alpha_s^2 \sin{2 \theta_{\sss L}} e^{i
\delta_{\sss L}}}{4 \sqrt{2} \gf m_{\tilde g}^2} \left[
-\frac{5}{9}F(x_{\tilde b_{\sss L} \tilde g},x_{\tilde
q_{\sss L} \tilde g}) +\frac{1}{36} G(x_{\tilde b_{\sss L}
\tilde g},x_{\tilde q_{\sss L} \tilde g}) + \frac{1}{2}
A(x_{\tilde b_{\sss L} \tilde g}) + \frac{2}{9} B(x_{\tilde
b_{\sss L} \tilde g})\right] \nn\\
& & ~~~~~~~~~~~~~~~~~~~~
- (\tilde b_{\sss L} \to \tilde s_{\sss L})~,\nn\\
c_4^q &=& \frac{\alpha_s^2 \sin{2 \theta_{\sss L}} e^{i
\delta_{\sss L}}}{4 \sqrt{2} \gf m_{\tilde g}^2} \left[
\frac{1}{3}F(x_{\tilde b_{\sss L} \tilde g},x_{\tilde q_{\sss
L} \tilde g}) +\frac{7}{12} G(x_{\tilde b_{\sss L} \tilde
g},x_{\tilde q_{\sss L} \tilde g}) - \frac{3}{2} A(x_{\tilde
b_{\sss L} \tilde g}) - \frac{2}{3} B(x_{\tilde b_{\sss L}
\tilde g})\right] \nn\\
& & ~~~~~~~~~~~~~~~~~~~~
- (\tilde b_{\sss L} \to \tilde s_{\sss L})~,\nn\\
c_5^q &=& c_6^q = 0~,
\eea
where $x_{ab} = m_a^2/m_b^2$.  Wilson coefficients with
inverse chirality $\tilde c$'s have exactly the same form,
with the replacement $L\leftrightarrow R$.  Loop integrals
are given by
\bea
F(x,y) &=& -\frac{x \ln{x}}{(x-y) (x-1)^2} -\frac{y
  \ln{y}}{(y-x) (y-1)^2} - \frac{1}{(x-1)(y-1)}~,\nn\\
G(x,y) &=& \frac{x^2 \ln{x}}{(x-y) (x-1)^2} +\frac{y^2
\ln{y}}{(y-x) (y-1)^2} + \frac{1}{(x-1)(y-1)}~,\nn\\
A(x) &=& \frac{1}{2(1-x)} + \frac{(1+2x) \ln{x}}{6
(1-x)^2}~,\nn\\
B(x) &=& -\frac{11-7x+2x^2}{18(1-x)^3} -
\frac{\ln{x}}{3(1-x)^4}~.
\eea

Finally, for the chromomagnetic penguin, we have
\beq
\lambda_t \frac{2 \alpha_s}{3 \pi} c_{8g}^{eff} = \frac{8}{3}
\frac{\alpha_s^2 \sin{(2 \theta_{\sss L})} e^{i \delta_{\sss
L}}}{4 \sqrt{2} \gf m^2_{\tilde g}} \left[ f_8^{\sss
{\rm SUSY}} (x_{\tilde b_{\sss L} \tilde g}) - (b_{\sss L}
\leftrightarrow s_{\sss L}) \right]~,
\eeq
where
\beq
f_8^{\sss {\rm SUSY}}(x) = \frac{-11+51 x
-21x^2-19x^3-6x(1-9x)\log{x}}{72(x-1)^4}
\eeq

The equations presented above allow one to calculate Wilson
coefficients at the SUSY scale, taken to be $m_t$.  They then
need to be renormalized to the scale $\mu = m_b$. The
renormalization procedure described in Ref.~\cite{BsKK} is
used. This then gives the three NP SUSY amplitudes
$\ApNPcomb$, $\ApNPCuph$ and $\ApNPCdph$ at scale $m_b$.

We can now see if GNK can explain the $\btopik$ puzzle.  In
Ref.~\cite{BtopiKfits}, fits were done with NP. The value of
$\gamma$ was taken from independent measurements. (The value
of $\gamma$ is the same as in the SM even in the presence of
NP \cite{patterns}.) However, if all NP amplitudes are kept,
there are more theoretical parameters (10) than measurements
(9), and a fit cannot be done.  For this reason, a single NP
amplitude was assumed to dominate. Four possibilities were
considered: (i) only ${\cal A}^{\prime, comb} \ne 0$, (ii)
only ${\cal A}^{\prime {\sss C}, u} \ne 0$, (iii) only ${\cal
A}^{\prime {\sss C}, d} \ne 0$, (iv) $\ApNPCuph = \ApNPCdph$,
${\cal A}^{\prime, comb} = 0$ (isospin-conserving NP). A very
good fit was found only if the NP is in the form of
$\ApNPcomb$ (i.e.\ the SM electroweak-penguin amplitude). It
is therefore often said that any NP invoked to explain the
$\btopik$ puzzle must contribute mainly to ${\cal A}^{\prime,
comb}$ and little to ${\cal A}^{\prime {\sss C}, u}$ and
${\cal A}^{\prime {\sss C}, d}$.  (However, it should be
noted that the fit with only ${\cal A}^{\prime {\sss C}, u}
\ne 0$ is not bad.) On the other hand, the GNK SUSY model
gives nonzero values to all three NP amplitudes, and so the
results of Ref.~\cite{BtopiKfits} do not hold. Another
procedure must be used.

Our analysis proceeds as follows. The three NP SUSY
amplitudes depend on a number of theoretical inputs.  We
generate these randomly in the following ranges:
\begin{itemize}
\item $300 \le m_{\tilde g} \le 2000$ GeV,
\item $100 \le m_{\tilde q} \le 2000$ GeV,
\item $-\pi/4 < \theta_{\sss L,R} < \pi/4$,
\item $-\pi < \delta_{\sss L,R} < \pi$,
\item $\gamma = {67.6^{+2.8}_{-4.5}}^\circ$ \cite{CKMfitter},
\item $m_u, m_d~(2~{\rm GeV}) = 2.5$ to $5.5$ MeV \cite{pdg},
\item $m_s~(2~{\rm GeV}) = 0.095 \pm 0.025$ GeV \cite{pdg},
\item $F^{\sss B \to K}(q^2=0) = 0.34 \pm 0.05$
\cite{BenekeNeubert},
\item $F^{\sss B \to \pi}(q^2=0) = 0.28 \pm 0.05$
\cite{BenekeNeubert}.
\end{itemize}
Note that we have taken $m_{\tilde u_L} = m_{\tilde d_L}$
following $SU(2)_{\sss L}$ symmetry.  The weak phase $\gamma$
is allowed to vary in the $\pm 2\sigma$ range.  For the other
(theoretical) quantities for which an error is given, we take
the range as $\pm 1\sigma$. With these values, $\ApNPcomb$,
$\ApNPCuph$ and $\ApNPCdph$ are generated.

\begin{table}[tbh]
\center
\begin{tabular}{cccc}
\hline
\hline
Mode & $BR[10^{-6}]$ & $A_{\sss CP}$ & $S_{\sss CP}$ \\
\hline
$B^+ \to \pi^+ K^0$ & $23.1 \pm 1.0$ & $0.009 \pm 0.025$ & \\
$B^+ \to \pi^0 K^+$ & $12.9 \pm 0.6$ & $0.050 \pm 0.025$ & \\
$\bd \to \pi^- K^+$ & $19.4 \pm 0.6$ & $-0.097 \pm 0.012$ &
\\
$\bd \to \pi^0 K^0$ & $9.9 \pm 0.6$ & $-0.14 \pm 0.11$ &
$0.38 \pm 0.19$ \\
\hline
\hline
\end{tabular}
\caption{Branching ratios, direct CP asymmetries $A_{\sss
CP}$, and mixing-induced CP asymmetry $S_{\sss CP}$ (if
applicable) for the four $\btopik$ decay modes. The data is
taken from Refs.~\cite{HFAG} and \cite{piKrefs}.}
\label{tab:data}
\end{table}

Given the knowledge of the three NP amplitudes and $\gamma$,
the $\btopik$ amplitudes [Eq.~(\ref{BpiKNPamps})] and
observables depend only on the two SM diagrams $P'$ and $T'$
(magnitudes and strong phases; $\pewp$ is related to $T'$).
We can therefore do a fit to see how well the $\btopik$ data
is reproduced. If the $\chi^2_{min}$ is acceptable, then we
can conclude that the GNK SUSY model explains the $\btopik$
puzzle. If not, then it does not.

In order to establish what constitutes an ``acceptable'' fit,
we take our cue from ordinary observables.  There, the
$2\sigma$ limit implies that 4.55\% of the points of a
Gaussian distribution lie outside this interval.  In this
spirit, we assume that the $\chi^2_{min}$ is acceptable if
the percentage of the parameter space which has a worse
$\chi^2_{min}$ is 4.55\%, i.e. $\chi^2_{min}$ is taken to be
$< 11.31$. (Note: in practice, there is no relation between
Gaussian and $\chi^2_{min}$ distributions.  We use the
information from the Gaussian distribution only as a guide.)

Before presenting the conclusions of this analysis, we must
consider other constraints. There are many constraints on
SUSY models -- electroweak precision tests, $\Delta m_d$,
$\Delta m_{\sss K}$, $b \to s (d) \gamma$, etc. However, by
far the most stringent is that coming from $\bs$-$\bsbar$
mixing.  This is discussed in detail in Ref.~\cite{Baek}, and
we closely follow the analysis presented here. We find that
$|\Delta m_s/ \Delta m_s^{\sss SM}| = 0.788 \pm 0.195$. This
limits the SUSY contribution to $\bs$-$\bsbar$ mixing. Using
the expression given in Ref.~\cite{Baek}, we compute the GNK
SUSY contribution to $|\Delta m_s|$.  To do so, three more
theoretical parameters are needed, and we generate them
randomly:
\begin{itemize}
\item $B_1 = 0.86 _{-0.04}^{+0.05}$ \cite{Baek},
\item $B_4 = 1.17 _{-0.07}^{+0.05} $ \cite{Baek},
\item $B_5 = 1.94 _{-0.08}^{+0.23}$ \cite{Baek}.
\end{itemize}
For each set of theoretical parameters generated, we check
whether the constraint is satisfied (within $\pm 2\sigma$).

The parameter space of GNK SUSY models is enormous -- there
are 12 SUSY parameters alone.  In order to do our best to
adequately sample this parameter space, 500,000 sets of
theoretical parameters were generated.  For each set, we
checked whether the $\btopik$ and the $\bs$-$\bsbar$ mixing
data were reproduced. The results are shown in
Table~\ref{table2}, for the cases where (i) only $LL$ mixing
is allowed, and (ii) both $LL$ and $RR$ mixings are
allowed. From this Table we see that the case with only $LL$
mixing is preferred by the $\bs$-$\bsbar$ mixing
data. However, neither mixing scenario can explain the
$\btopik$ puzzle -- in both cases, the $\btopik$ data is
reproduced only in a tiny region of parameter space.  The
$\bs$-$\bsbar$ mixing constraint reduces this (already small)
region. We therefore conclude that it is very unlikely that
the GNK SUSY model obeys the constraints from $\btopik$
decays, and virtually impossible that it reproduces the data
from both $\btopik$ and $\bs$-$\bsbar$ mixing.

\begin{table*}[btc]
\center
\begin{tabular}{ccc}
\hline
\hline
$\chi^2_{min} < 11.31$ & $\Delta m_s$ & both \\
\hline
74 & 414357 & 15 \\ \hline
\hline
\end{tabular}
~~~~~~~~~~~\begin{tabular}{ccc}
\hline
\hline
$\chi^2_{min} < 11.31$ & $\Delta m_s$ & both \\
\hline
102 & 92844 & 1 \\ \hline
\hline
\end{tabular}
\caption{The number of points (out of 500000) which satisfy
$\chi^2_{min}(\btopik) < 11.31$, the $\Delta m_s$ constraint
within $\pm 2\sigma$, and both constraints. In the left
table, only $LL$ mixing is allowed, while in the right table,
both $LL$ and $RR$ mixings are allowed.}
\label{table2}
\end{table*}

\begin{figure}
	\centering
		\includegraphics[height=4.75cm]{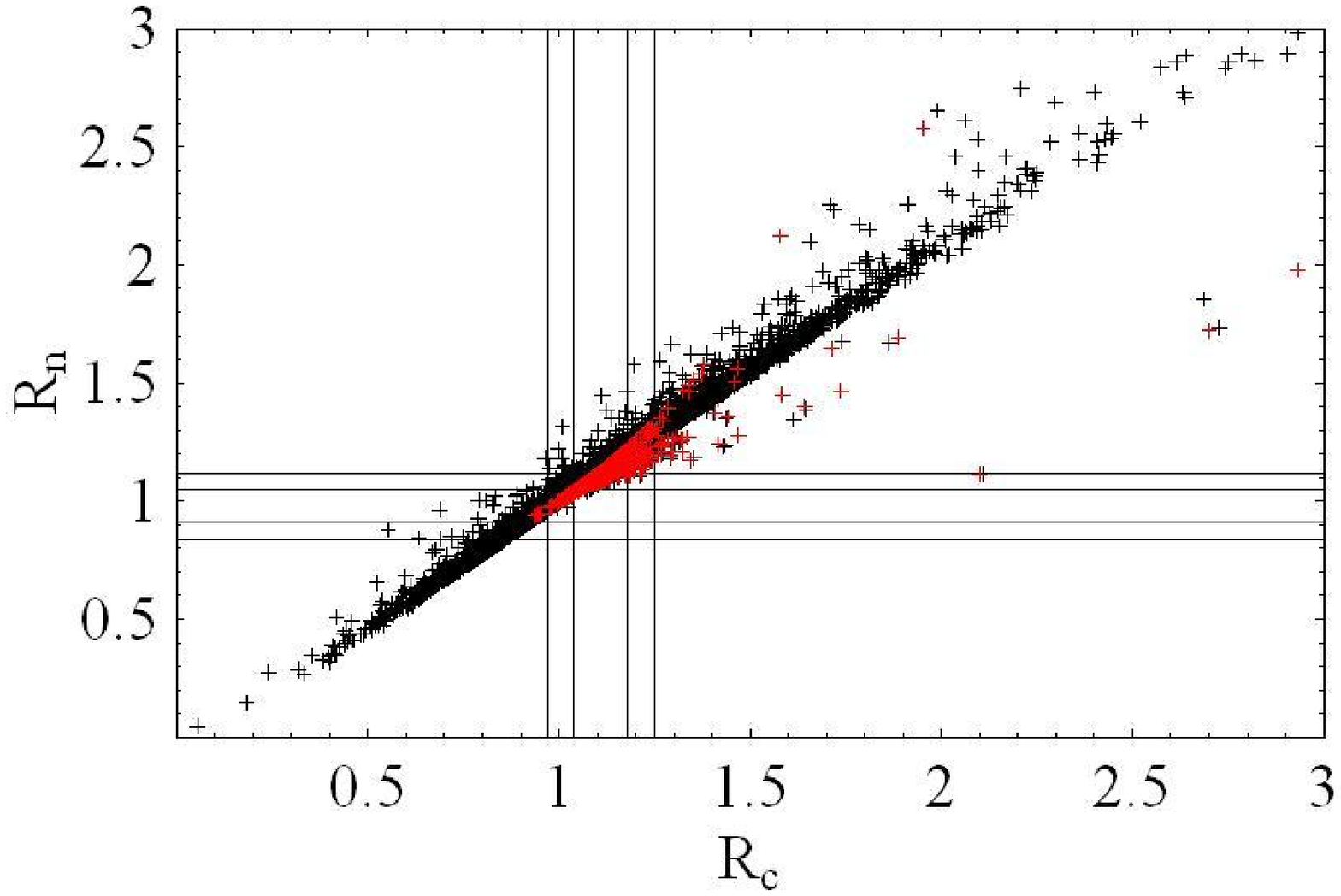}
		\includegraphics[height=4.75cm]{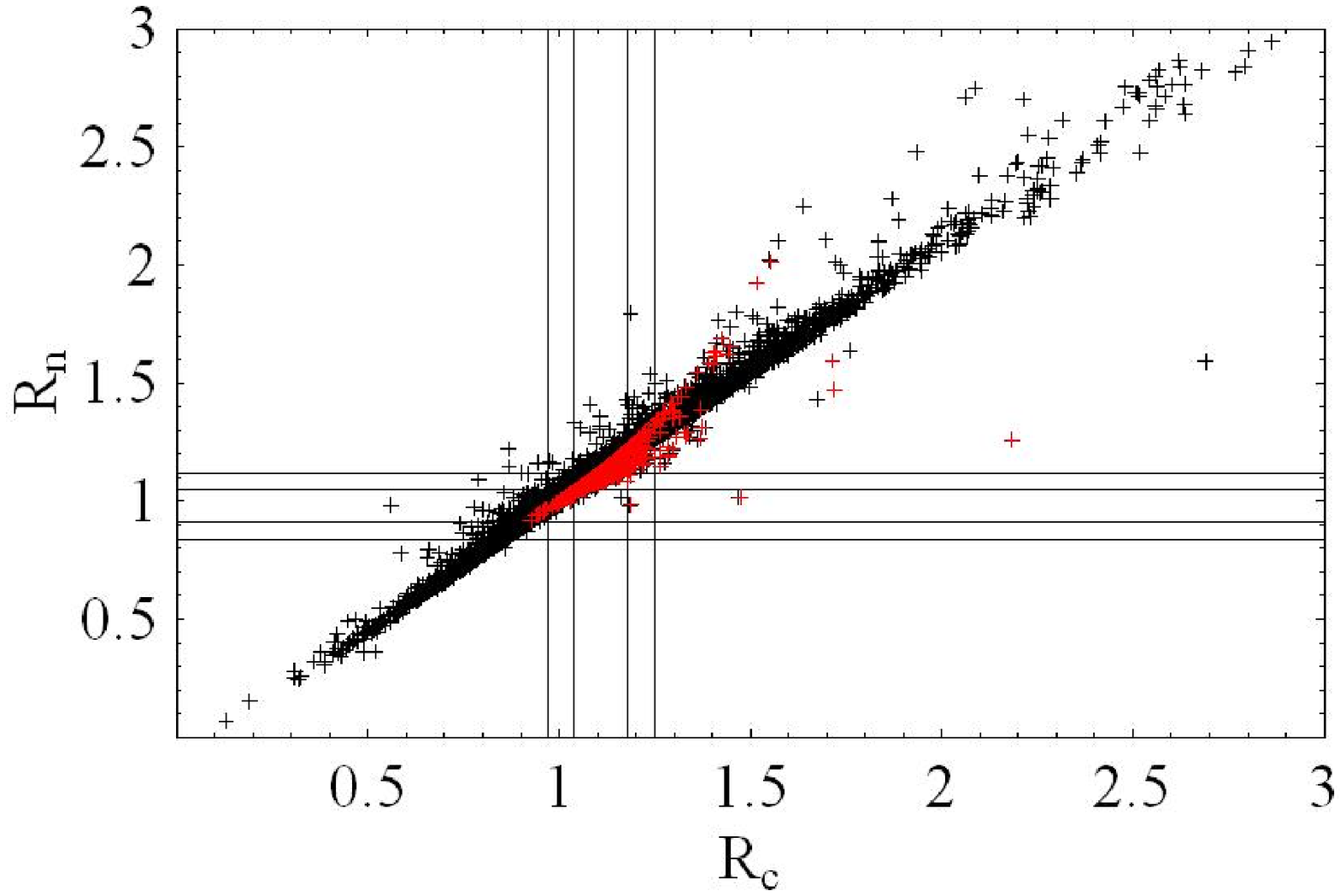}
	\centering
		\includegraphics[height=4.75cm]{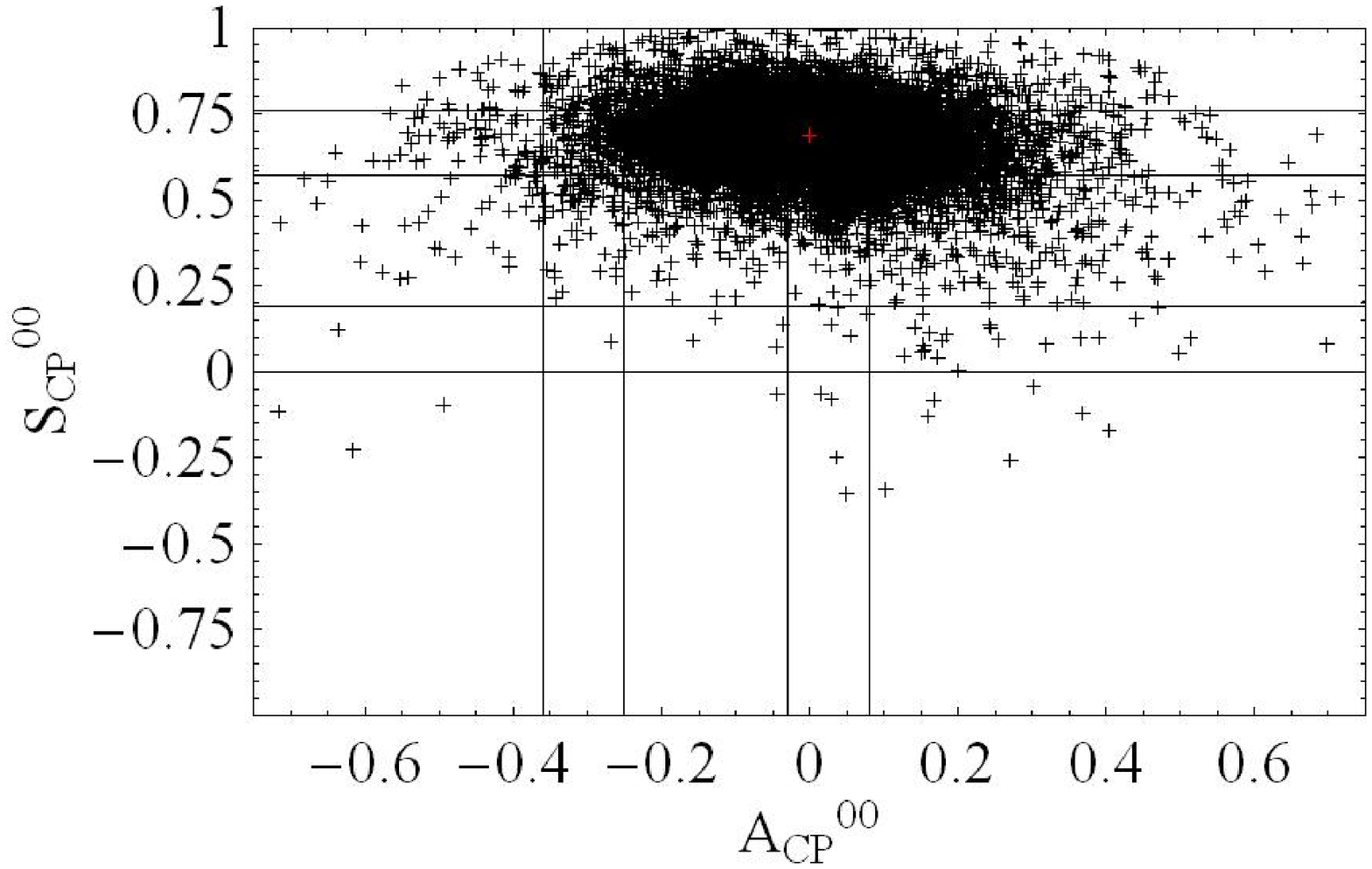}
		\includegraphics[height=4.75cm]{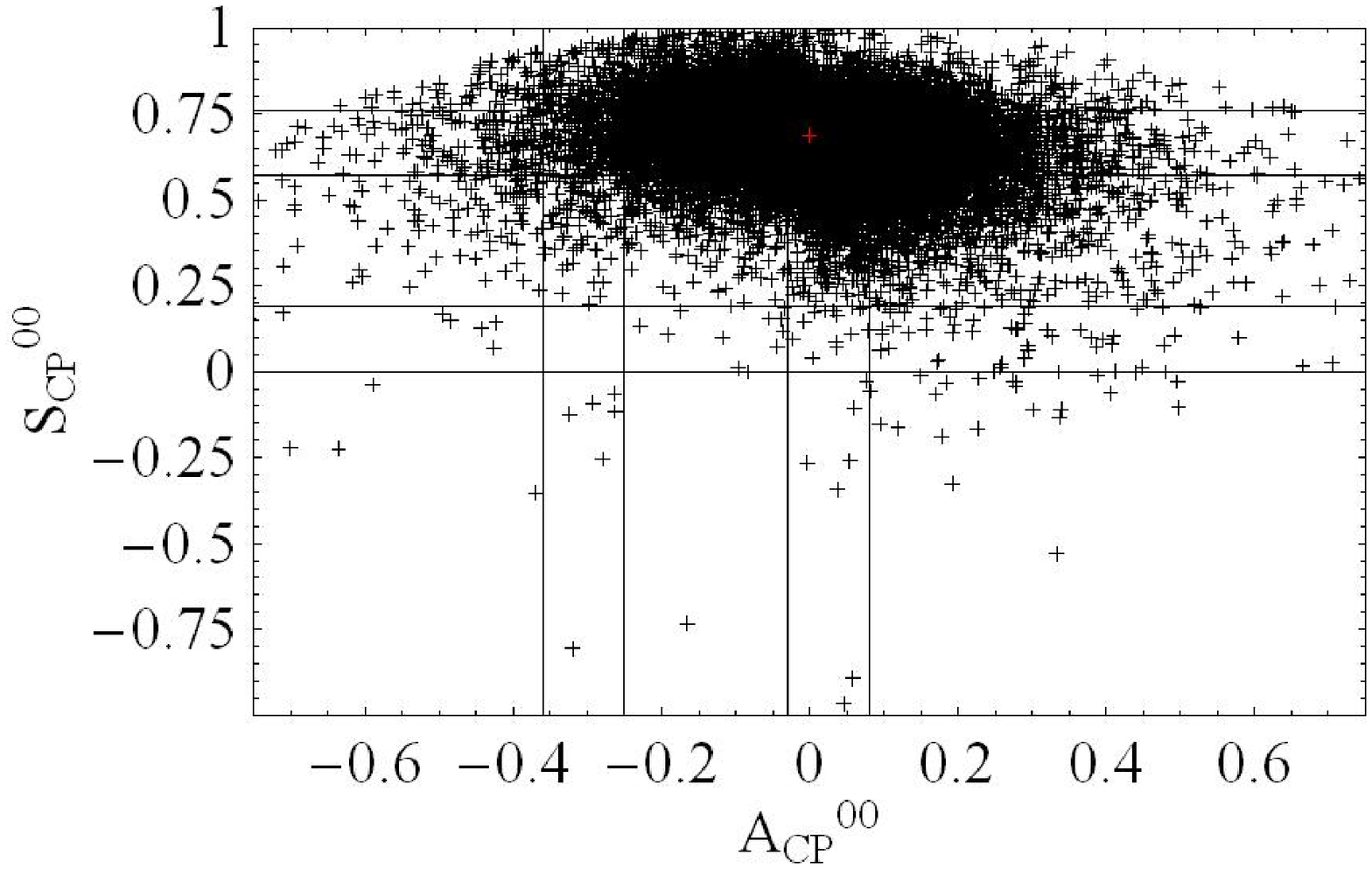}
	\centering
		\includegraphics[height=4.75cm]{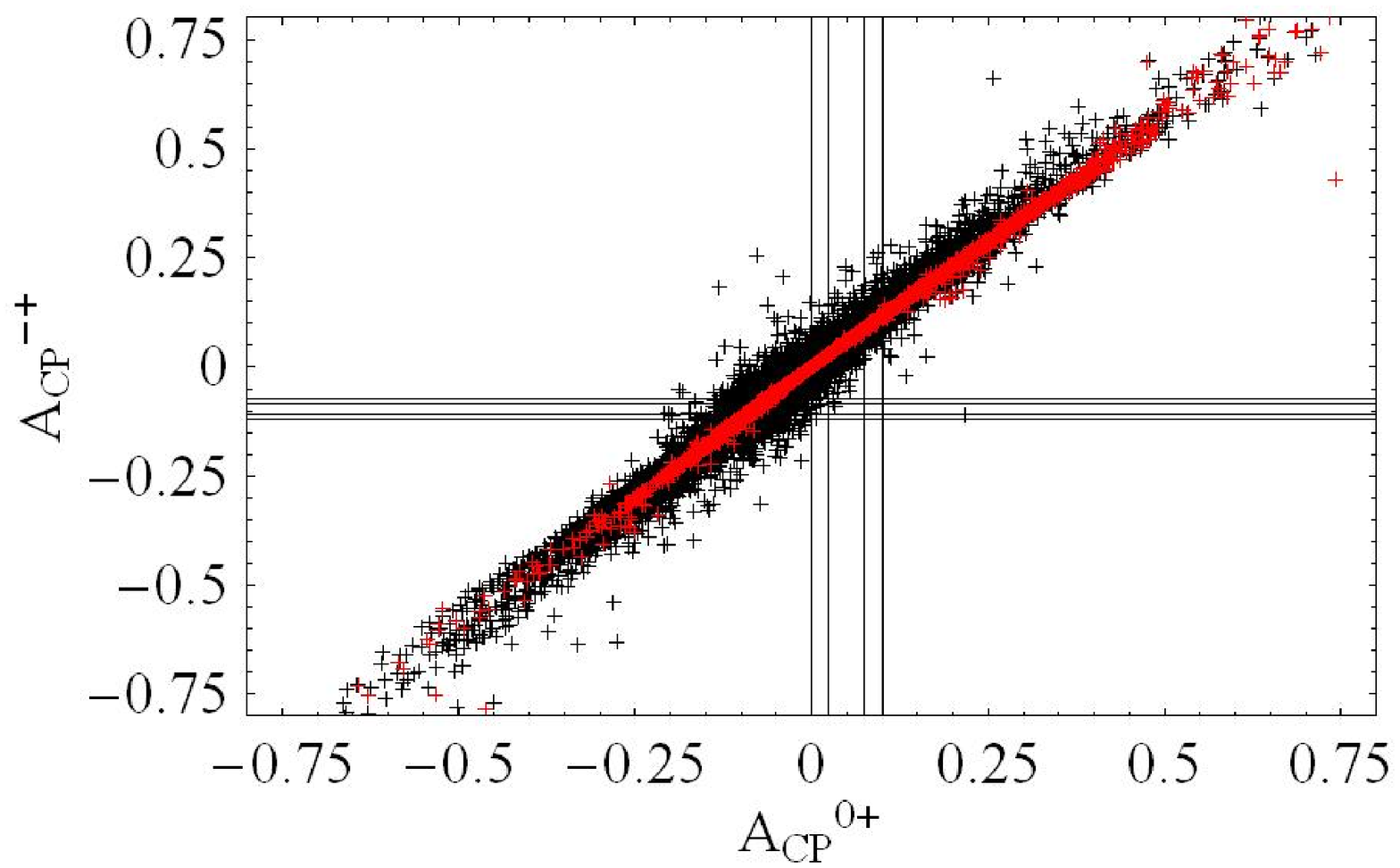}
		\includegraphics[height=4.75cm]{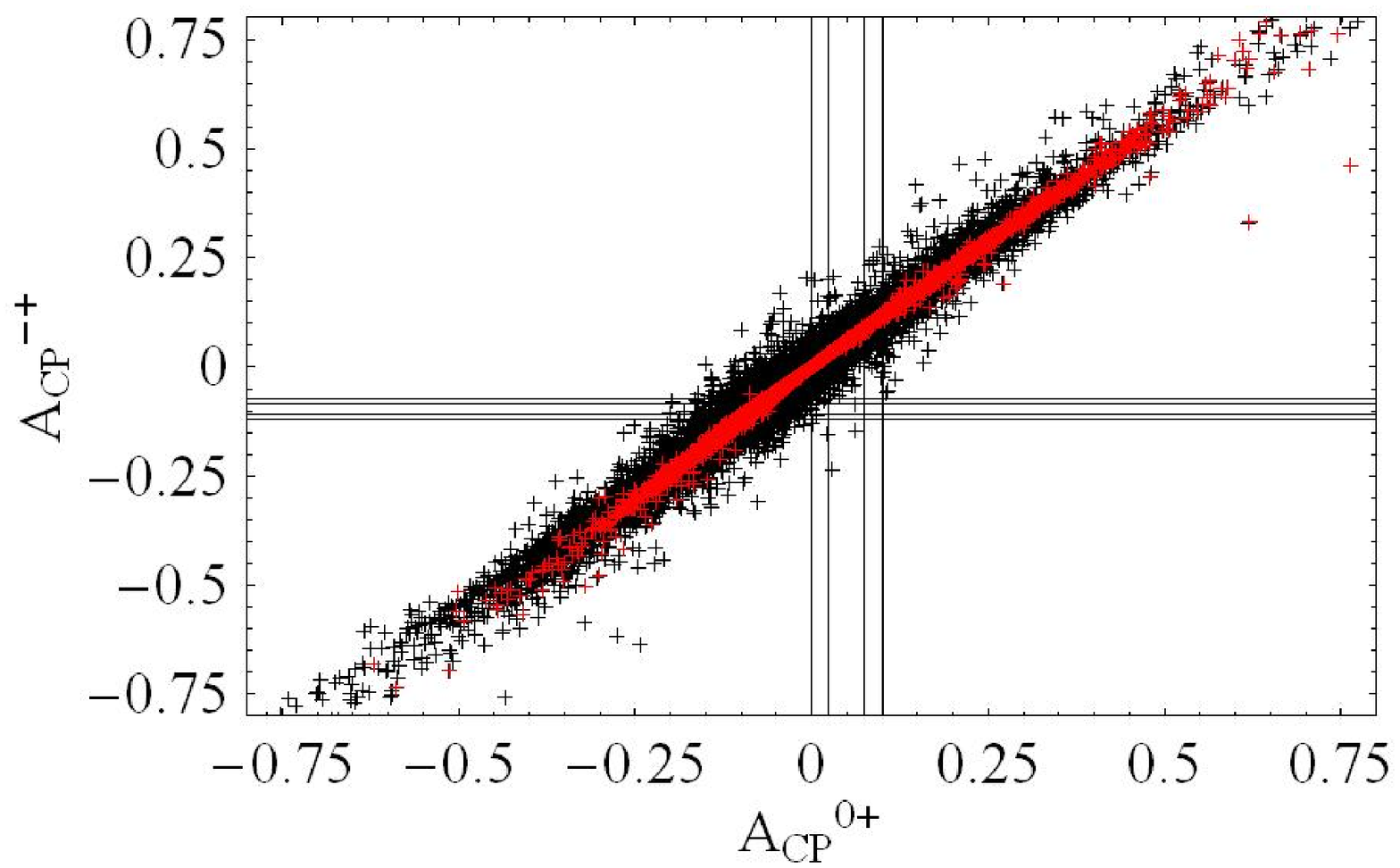}
\caption{Scatter plots of $R_c$ vs.\ $R_n$ (top),
$A^{00}_{\sss CP}$ vs.\ $S^{00}_{\sss CP}$ (middle), and
$A^{0+}_{\sss CP}$ vs.\ $A^{-+}_{\sss CP}$ (bottom), for LL
mixing only (left), and LL and RR mixing (right).  Horizontal
and vertical lines represent experimental values within
1$\sigma$ and 2$\sigma$. Plots include all 500,000 GNK sets
of parameters. Red points (dark grey in black and white)
indicate only the SM piece of the SM $+$ SUSY contribution.
For $A^{00}_{\sss CP}$ vs.\ $S^{00}_{\sss CP}$, the SM piece
is a single dot because there is no direct CP violation when
SUSY is not added.}
\end{figure}

In Fig.~1, we present the SUSY contributions to several
$\btopik$ observables.  This helps identify which
measurements lead to the large $\chi^2_{min}$ for each of the
500,000 GNK sets of parameters. In particular, we show $R_c$
vs.\ $R_n$, where
\bea
R_c &\equiv& 2\left[\frac{\mbox{BR}(B^+\to\pi^0K^+)+
\mbox{BR}(B^-\to\pi^0K^-)}{\mbox{BR}(B^+\to\pi^+ K^0)+
\mbox{BR}(B^-\to\pi^- \bar K^0)}\right] ~,\nn\\
R_n &\equiv& \frac{1}{2}\left[\frac{\mbox{BR}(\bd\to\pi^-
K^+)+ \mbox{BR}(\bdbar\to\pi^+ K^-)}
{\mbox{BR}(\bd\to\pi^0K^0)+ \mbox{BR}(
\bdbar\to\pi^0\bar K^0)}\right] ~,
\eea
$A^{00}_{\sss CP}$ vs.\ $S^{00}_{\sss CP}$, and $A^{0+}_{\sss
CP}$ vs.\ $A^{-+}_{\sss CP}$.  All are scatter plots, showing
the contribution of the GNK SUSY model to the various
observables.  As can be seen from this Figure, GNK has little
difficulty in reproducing the combined $R_c$ and $R_n$
quantities. However, the SM can do this alone, showing that
there is no discrepancy with the SM for $R_c$ and $R_n$. GNK
can also explain the $A^{00}_{\sss CP}$ and $S^{00}_{\sss
CP}$ observables.  Note that the SM alone has difficulty with
these measurements. On the other hand, it is almost
impossible for GNK to simultaneously reproduce $A^{0+}_{\sss
CP}$ and $A^{-+}_{\sss CP}$.  This shows explicitly that it
is the direct CP asymmetry measurements which are most
problematic.

There are several reasons that the GNK SUSY model cannot
explain the $\btopik$ puzzle.  First, for much of the
parameter space, all three NP amplitudes are small.  Thus,
despite the presence of SUSY, the $\btopik$ system is
basically described by the SM.  However, we saw that the SM
has a very poor fit in explaining the $\btopik$ observables,
and so the same is true here. Second, the $\btopik$
measurements suggest that there is NP in the $\pewp$ diagram
($\ApNPcom$).  However, as indicated earlier, SUSY does not
contribute significantly to $\pewp$.  As a result, it is very
difficult for SUSY to explain the $\btopik$ puzzle, and the
GNK SUSY fits are generally poor.  Third, we saw in the fits
in which a single NP amplitude was assumed to dominate that
the fit with ${\cal A}^{\prime {\sss C}, u} \ne 0$ was not
bad.  However, GNK generally does not generate only a large
${\cal A}^{\prime {\sss C}, u}$ -- a large ${\cal A}^{\prime
{\sss C}, d}$ is also usually found.  Again, this leads to a
poor fit. All of these can be seen in Fig.~2, which shows the
plots of ${\cal A}^{\prime {\sss C}, u}$ vs.\ $\ApNPcom$ and
${\cal A}^{\prime {\sss C}, u}$ vs.\ ${\cal A}^{\prime {\sss
C}, d}$. The bottom line is that it requires a very precise
pattern of SUSY parameters to explain the $\btopik$ puzzle,
and this is not found in most of the GNK SUSY parameter
space.

\begin{figure}
\centering
	\includegraphics[height=4.75cm]{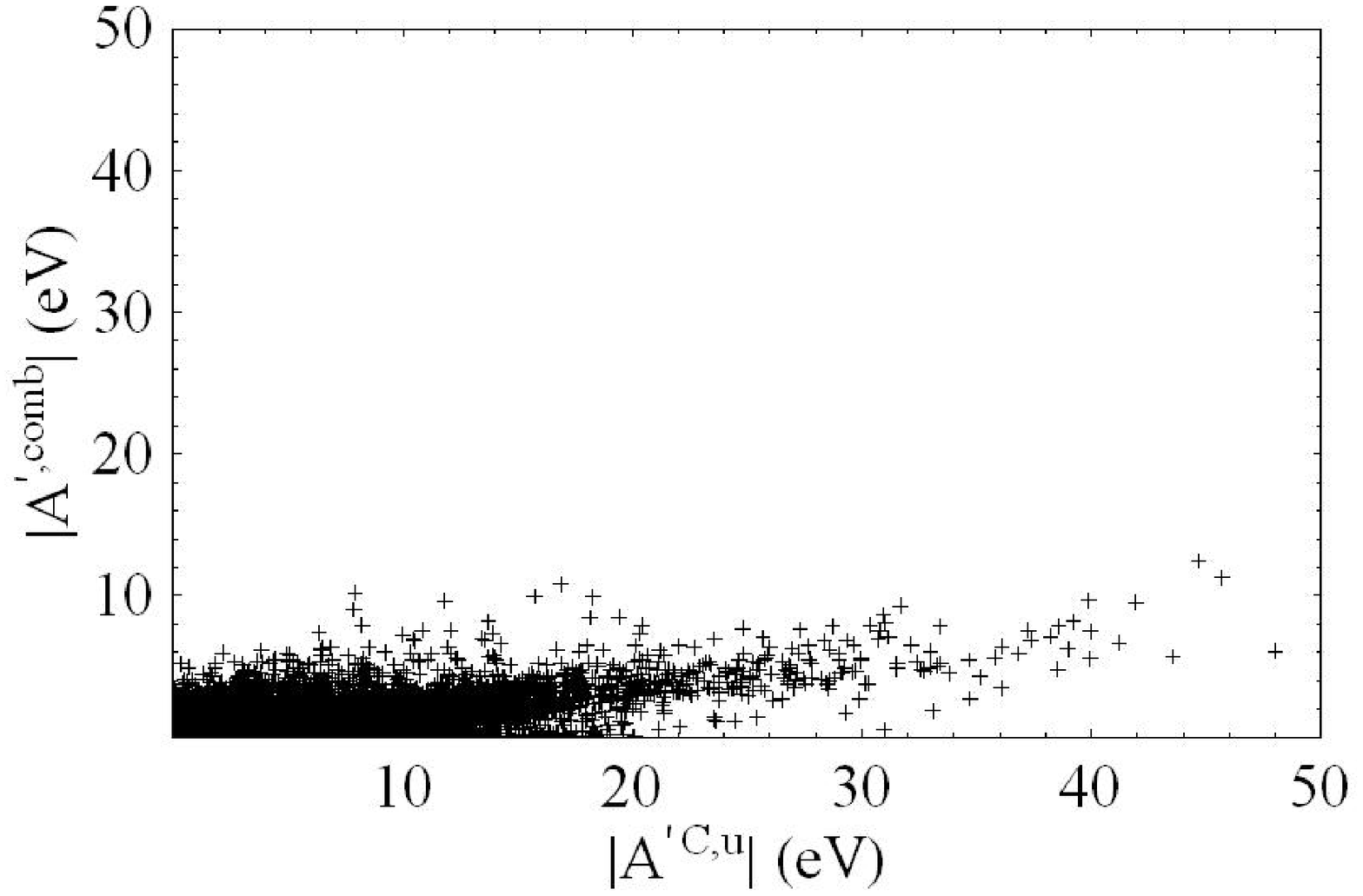}
	\includegraphics[height=4.75cm]{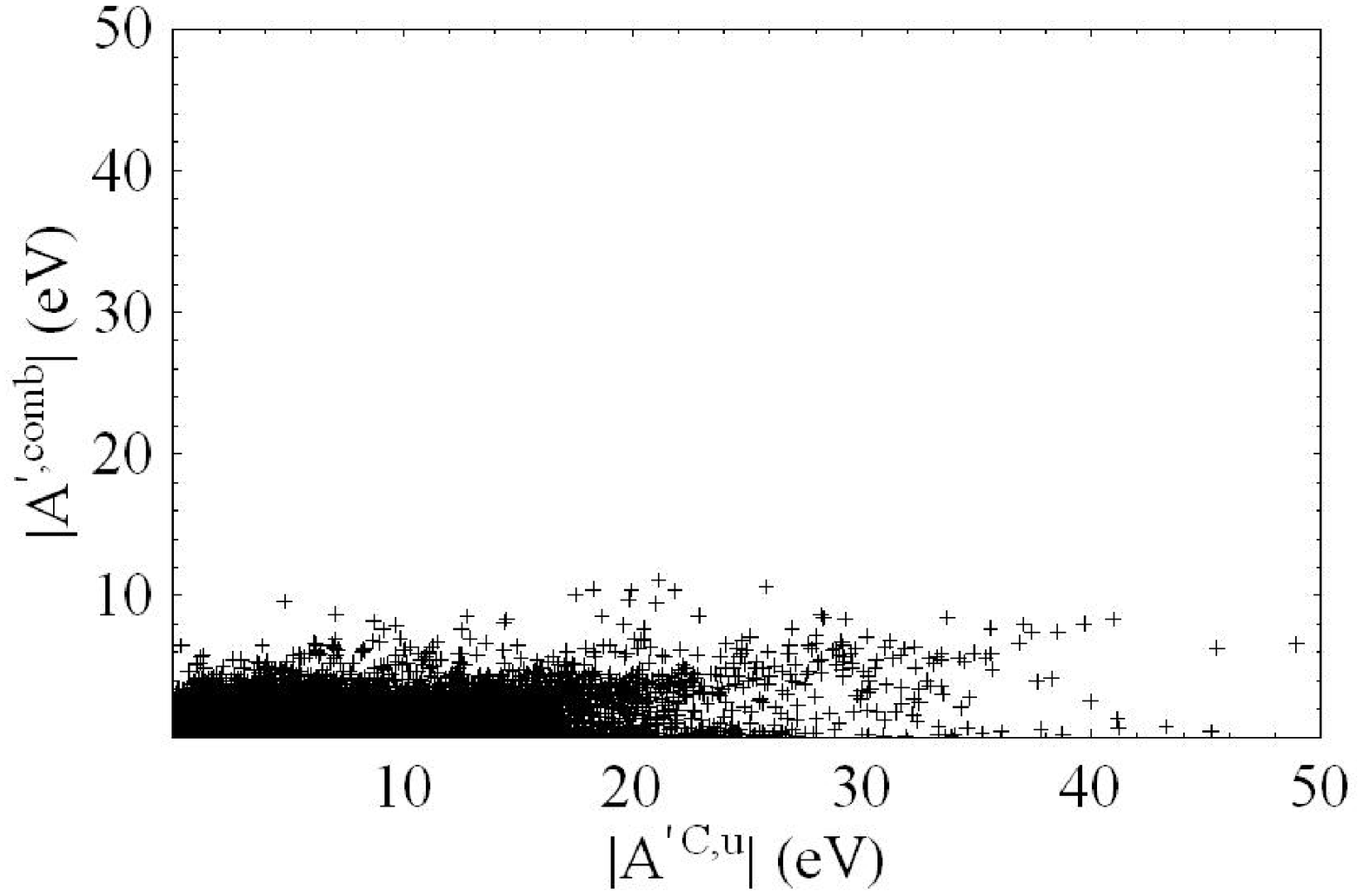}
\centering
	\includegraphics[height=4.75cm]{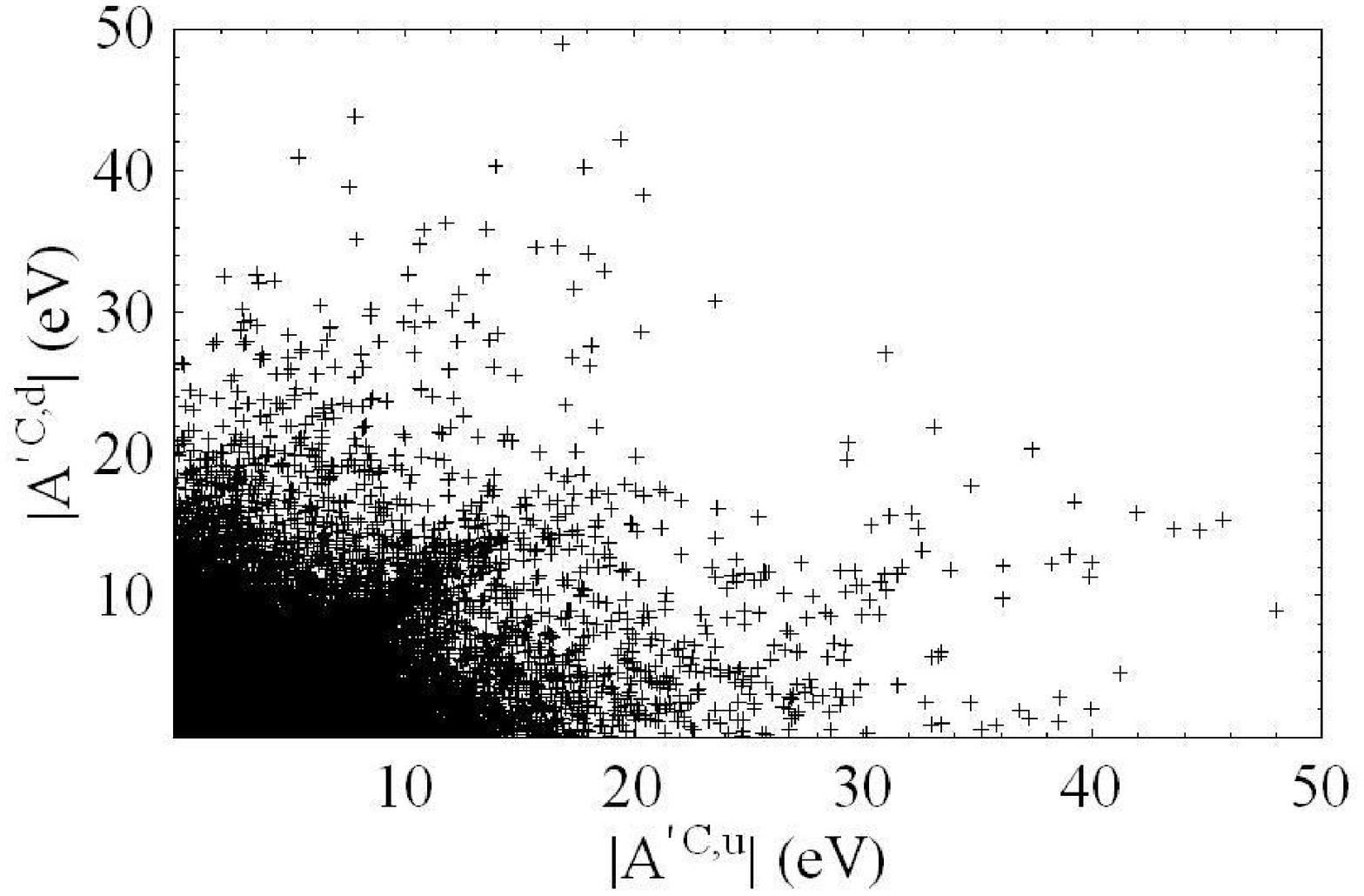}
	\includegraphics[height=4.75cm]{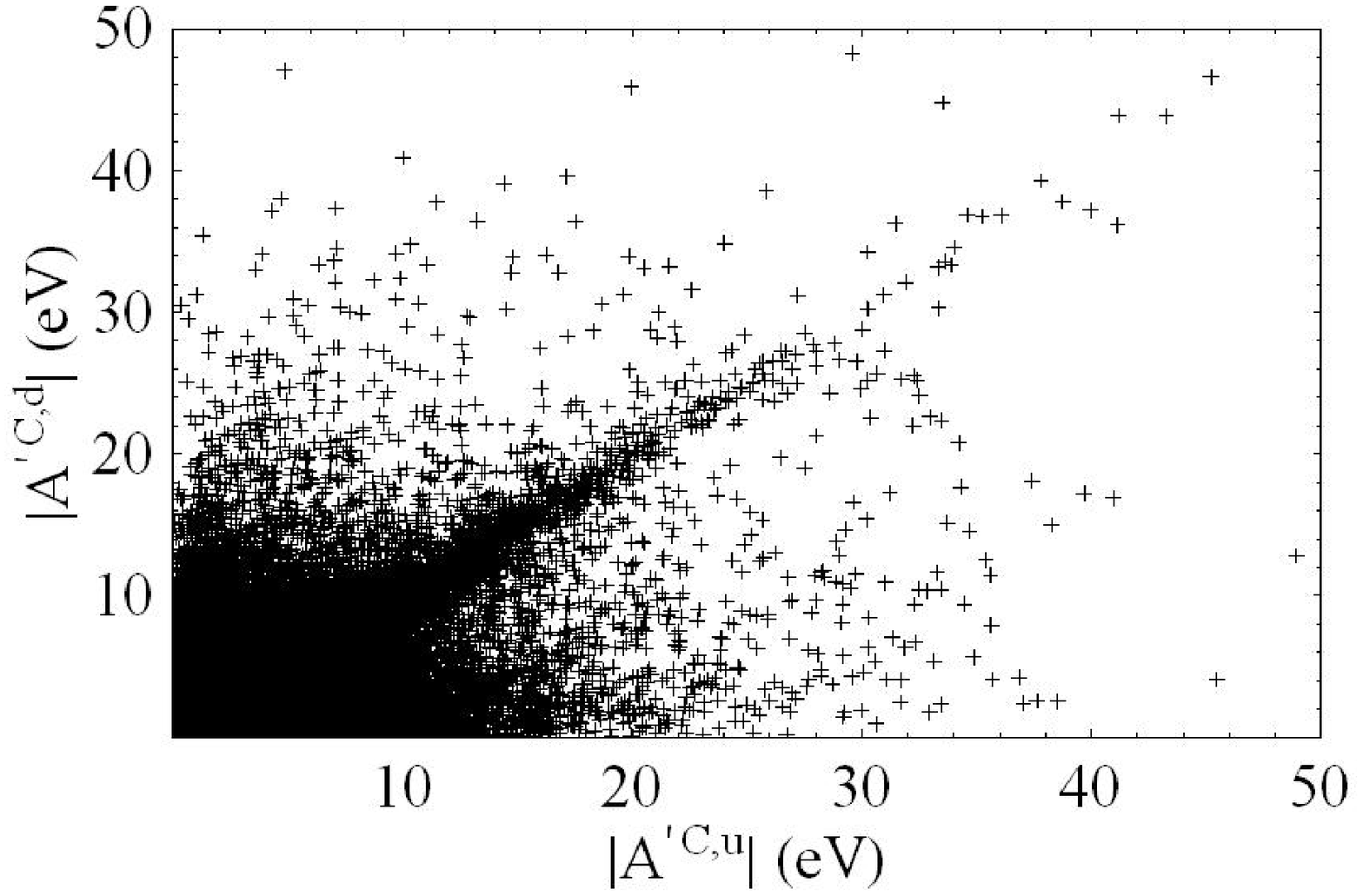}
\caption{Scatter plots of ${\cal A}^{\prime {\sss C}, u}$
  vs.\ $\ApNPcom$ (top) and ${\cal A}^{\prime {\sss C}, u}$
  vs.\ ${\cal A}^{\prime {\sss C}, d}$ (bottom), for LL
  mixing only (left), and LL and RR mixing (right). Plots
  include all 500,000 GNK sets of parameters.}
\end{figure}

Of the very few points which satisfy both constraints, the
great majority correspond to a large ${\cal A}^{\prime {\sss
C}, u}$ and a small ${\cal A}^{\prime {\sss C}, d}$ and
$\ApNPcom$.  Also, all the points with
$\chi^2_{min}(\btopik) < 11.31$ have a gluino mass less than
1.3 TeV.  This is the only direct constraint on the SUSY
parameters.

As we have seen, it is extremely unlikely that the GNK SUSY
model explains the $\btopik$ puzzle. As noted earlier, there
are other popular SUSY models: mSUGRA \cite{mSUGRA}, AMSB
\cite{AMSB}, GMSB \cite{GMSB}, etc, However they all
automatically solve the SUSY FCNC/CP problems by not allowing
any CP-violating phases. So these models cannot explain the
$\btopik$ data either.

There are two SUSY models which do reproduce the $\btopik$
data. They have (i) a large chargino contribution which
allows large (2,3) mass terms in the up-squark sector
\cite{Khalil}, or (ii) R-parity violation \cite{Yang}.
However, these two models have their own problems. The one
with chargino contributions seems to be fine-tuned. It is not
natural, i.e.\ it is hard to find a more microscopic theory
which generates only (2,3) up-squark mass components in the
$LL$ or $RR$ sector. And the R-parity-violating model lacks
the beauty of SUSY, e.g.\ it does not have dark-matter
candidates. We therefore conclude that if the $\btopik$
puzzle persists, SUSY models could have some difficulty.

To summarize, the supersymmetry (SUSY) model of Grossman,
Neubert and Kagan (GNK) \cite{GNK} has great difficulty in
explaining the $\btopik$ puzzle.  The $\btopik$ data can be
reproduced in the GNK model, but only in a tiny region of
parameter space.  Other SUSY models, such as those with
minimal supergravity \cite{mSUGRA}, anomaly-mediated SUSY
breaking \cite{AMSB} or gauge-mediated SUSY breaking
\cite{GMSB}, fare no better, as they do not allow any new
CP-violating phases. There are two SUSY models which do
reproduce the $\btopik$ data \cite{Khalil,Yang}. However,
these models are either fine-tuned or lack some elements of
ordinary SUSY theories. The $\btopik$ puzzle is still only a
$\gsim 3\sigma$ effect, and so cannot be considered
statistically significant.  However, if this discrepancy with
the SM remains in the years to come, it could pose a problem
for SUSY models.

\bigskip
\noindent {\bf Acknowledgments}:

This work is financially supported by NSERC of Canada (MI and
DL) and by the Korea Research Foundation Grant
funded by the Korean Government (MOEHRD) No. KRF-2007-359-C00009 (SB).


\end{document}